\documentstyle[twoside,epsfig]{hsproc}

\def\runauthor{J.~Rak for PHENIX Collaboration}
\def\shorttitle{Azimuthal Correlations at RHIC}

\def\bgi{\begin{itemize}}
\def\endi{\end{itemize}}
\def\bge{\begin{equation}}
\def\ende{\end{equation}}
\def\bgc{\begin{center}}
\def\endc{\end{center}}

\def \pt{$p_{\perp}$}
\def \vvpt{$v_{2}(p_\perp)$}
\def \vv{$v_{2}$}
\def \mcos2{$\cos{(2\Delta\phi)}$}
\def \gev   {~GeV}
\def \gevc  {~GeV/$c$}

\def \roots {$\sqrt{s_{NN}}$}

\def	\vpt	{$\vec{p}_{\perp}$}
\def	\vjT	{$\vec{\j}_{\perp}$}
\def	\jT	{$j_{\perp}$}

\def	\mjT	{$\sqrt{\langle j_{\perp}^2\rangle}$}
\def	\mkT	{$\sqrt{\langle k_{\perp}^2\rangle}$}
\def	\mjTy	{$\langle |j_{\perp y}|\rangle$}
\def	\mkTy	{$\langle |k_{\perp y}|\rangle$}
\def	\mpT	{$\langle p_{\perp}\rangle$}

\begin{document}

\title{High-$p_\perp$ charged particle azimuthal correlations at RHIC}

\author{Jan Rak\email{janrak@bnl.gov} for the PHENIX collaboration}{Iowa State University}

\abstract{
The analysis of two-particle azimuthal correlation function in $p+p$
and $Au+Au$ collisions at \roots = 200 \gev\ is presented. The mean
jet fragmentation transverse momentum \mjT\ and mean parton transverse
momentum \mkT\ are inferred from $p+p$ collisions. The analysis of
$Au+Au$ data, where elliptic flow is a dominant source of
two-particles correlation is also presented. We derive the upper limit
on partonic azimuthal anisotropy based on jet-quenching in a Glauber
model. We show that some new mechanism, generating the azimuthal
anisotropy observed in final state, is needed.
}

\section{Introduction}

With the beginning of RHIC operation heavy-ion physics entered a new
regime, where pQCD phenomena can be fully explored. High-energy
partons materializing into hadronic jets can be used as sensitive
probes of the early stage of heavy ion collisions.
Measurements carried out during the first two years of RHIC operation
at \roots=130 and 200 GeV exhibit many surprising features. The
high-\pt\ particle yield was found to be strongly suppressed in $Au+Au$
central collisions \cite{ppg001,ppg014}, and the particle
distribution in azimuthal space reveals large asymmetries attributed
to sizable elliptic flow \cite{ppg004}. Furthermore, the azimuthal
anisotropy in the region of \pt$>$2.5-3\gevc, where the contribution
from hydrodynamics is expected to be negligible, is found to be
significant and \pt\ independent \cite{STARv2}. Another
unique feature, found in RHIC data, is the observation of back-to-back
jet correlation disappearance in central $Au+Au$ collisions
\cite{STARb2b}.  Although tremendous progress in the understanding
of these phenomena has been made, an unambiguous picture merging all
observables in a consistent way still has to be developed.

The QCD medium effect on high-\pt\ particle production in heavy ion
collisions is quantified with the {\sl nuclear modification factor}
$R_{AA}(p_\perp)$ given by the ratio of the measured $AA$ invariant
yields to the $NN$ collision scaled $p+p$ invariant yields (eq. (1) in 
\cite{ppg014}). With no nuclear modification of particle yield, $R_{AA}(p_\perp)=1$.
The large decrease of $R_{AA}(p_\perp)$ observed in central $Au+Au$ collisions is
commonly attributed to the multiple interactions of highly energetic
quarks with excited nuclear medium. Induced soft-gluon radiation
carries off a significant fraction of each parton's energy, which leads to
the depletion of the high-\pt\ particle yield \cite{GyulassyVitev}.

It is expected that the same mechanism of parton energy loss in the
nuclear medium should also generate azimuthal anisotropies observed in
the high-\pt\ region. However, as it was first pointed out by E.Shuryak
\cite{Shuryak} and B. Muller \cite{Muller}, the influence of the
nuclear medium on parton trajectory, although sufficient for excessive
yield suppression, is not capable to generate azimuthal anisotropies
larger than 10\%. 

It has been suggested that the partonic ``collective'' hadronization
might amplify the hadronic flow to the values of \vv, as seen in
experiment, if one assumes some sort of partonic coalescence
\cite{coal1,coal2,coal3}. The \vv\ values of identified $\pi^\pm$,
$K^\pm$ and $p\bar{p}$ \cite{ppg022}, and $K^0_S$,
$\Lambda\bar{\Lambda}$ \cite{STAR_ident_v2} scaled by its quark
content are apparently suggesting some kind of partonic collectivity.
However, more data, especially identified particles at high-\pt, are
needed.
 
Relatively good quantitative description of the high-\pt\ \vvpt\ data
was achieved in the model \cite{Kovchegov02,Kovchegov03} based on
extension of the original ``Color Glass Condensate'' ($CGC$) model
\cite{larry,dima}.  The mono-jet (one gluon) radiation in classical
$CGC$ has been extended by including two-gluon production
amplitude. This two-gluon radiation term generates two-particle
azimuthal correlations, which are evidently strong enough to account
for the measured strength of \vvpt. The elliptic flow in this model
does not respect the reaction plane, and the azimuthal anisotropy is
solely of the two-particle correlations nature. However, the absence of the
correlation between spatial distribution of particle emission and the
orientation of the reaction plane in this model seems to contradict some of
the other observation, e.g. reaction plane dependent HBT
\cite{STAR_HBT}. Moreover, the recent results from $d+Au$ analysis
\cite{dAu} leave only small room for initial state phenomena required
by $CGC$ model.

In the next section we will present the analysis of the two-particle
correlation function in $p+p$ collisions, where the elliptic flow
correlation is not present and the jet properties can be studied in
a ``clean'' environment.

\section{Jet fragmentation in $p+p$ collisions \\ at \roots=200\gev}

\begin{figure}[h]
\bgc
\includegraphics[width=8cm,height=3.5cm,
bbllx=26,bblly=190,bburx=550,bbury=420,clip=]{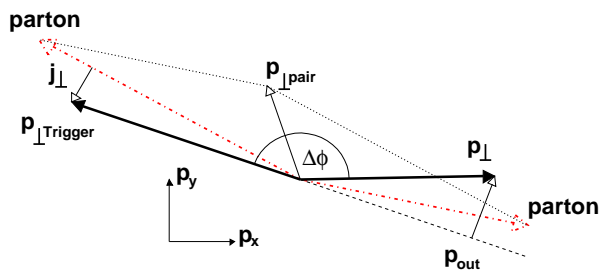}
\endc
\caption{
The schematic view of the hard scattering event. Two back-to-back
scattered partons fragment producing high-\pt\ particles labeled as
$p_{\perp Trigger}$ and \pt, \jT\ denotes the jet fragmentation
transverse momentum and $p_{\perp pair}$ the parton pair's net transverse momentum. }
\label{FigJets}
\end{figure}

Jets are produced in the hard scattering of two partons. The schematic
view of such an event in the plane perpendicular to the beam axis is
shown on Fig.~\ref{FigJets}.  Two scattered partons propagate
back-to-back from the collision point and fragment into the jet-like
spray of final state particles (only one fragment of each parton is
shown). These particles have a transverse momentum \vjT\ with respect
to the partonic transverse momentum. The magnitude of \vjT\ has been
observed to be \pt\ independent (see \cite{CCORjt}).

In the case of collinear  partonic collisions the two emerging partons
would have the same magnitude of transverse momenta pointing to the
opposite directions. However, the partons are carrying  initial intrinsic
momenta $\vec{k}_\perp$ before the collision happens.
This affects the outgoing direction and magnitude of partonic \pt.
It results in a momentum imbalance (the partons' \pt\ are not equal) and 
an acoplanarity (transverse momentum of one jet does not lie in the plane
determined by the transverse momentum of the second jet and the beam axes).
The jets are not collinear and have a net transverse momentum
$\sqrt{\langle p_{\perp}^2\rangle}_{pair} =
\sqrt{2}\cdot\sqrt{\langle k_{\perp}^2\rangle}$ of magnitude \cite{kt_E609_Apana}.
If the jet axes were known, then the average projection of \vjT\ into
the plane perpendicular to the beam axes can be determined in the following way
\bge
\langle |j_{\perp y}|\rangle = \langle p_{\perp}\rangle \sin{\langle|\phi_i-\phi_{jet}|\rangle}
\ende
where \mpT\ is the average \pt\ of the particles produced in the
fragmentation process. Here $\phi_{jet}$ and $\phi_i$ are azimuthal angles
of the parton and $i$-th fragment respectively.  However, we can only
measure the relative angular dispersion between two jet fragments $i$
and $j$ which is $\sigma_{N} \equiv \sqrt{ \langle(\phi_i-\phi_j)^2\rangle } =
\sqrt{2} \sqrt{ \langle\ (\phi_i-\phi_{jet})^2\rangle } =
\sqrt{\pi}\cdot \langle|\phi_i-\phi_{jet}|\rangle$. 
The final formula for \mjTy\ is

\bge\label{eq_jt}
\langle |j_{\perp y}|\rangle = 
\langle p_{\perp}\rangle\cdot\sin{{\sigma_N\over\sqrt{\pi}}}
\ende

In order to extract \mkT\ from measured width of away side
correlation peak, $\sigma_F$, we used the equation for the average
transverse momentum $\langle |p_{out}| \rangle$ out of the plane
defined by $\vec{p}_{Trigger}$ and the beam axis (see \cite{CCORjt}).
\bge\label{eq_pout}
\langle|p_{out}|\rangle^2 = \langle|j_{\perp y}|\rangle^2 + x^2_E
(\langle|j_{\perp y}|\rangle^2+2\langle|k_{\perp y}|\rangle^2)
\ende
where
\bge
x_E \equiv -{\vec{p}_{\perp}\cdot\vec{p}_{\perp Trigger}\over
|\vec{p}_{\perp Trigger}|^2}
\ende
with $\vec{p}_{\perp}$ being the momentum of the particles produced in
the opposite direction to the $\vec{p}_{\perp Trigger}$. For the case where 
the magnitudes of $\vec{p}_{\perp Trigger}$ and $\vec{p}_{\perp}$ are
similar, then
\bge
x_E \approx  -\cos(\langle|\Delta\phi|\rangle) = -\cos(\sqrt{2\over\pi}\sigma_{F})
\ende
The magnitude of $\langle |p_{out}| \rangle$ is expressed as 
\bge
\langle |p_{out}|\rangle =
\langle p_{\perp}\rangle\sin(\langle|\Delta\phi|\rangle) =
\langle p_{\perp}\rangle\sin{(\sqrt{2\over\pi}\sigma_{F})}
\ende
From (\ref{eq_pout}) and using (\ref{eq_jt}) one
can derive the relation
\bge\label{eq_kt}
\langle|k_{\perp y}|\rangle =
\langle p_{\perp}\rangle\cos{\sigma_N\over\sqrt{\pi}}
\sqrt{ {1\over2}\tan^2\left(\sqrt{2\over\pi}\sigma_F\right) -
\tan^2\left(\sqrt{1\over\pi}\sigma_N\right)}
\ende
In the limit of small angles, $\sigma_N\rightarrow 0$,
$\sigma_F\rightarrow 0$ and equations (\ref{eq_jt}) and (\ref{eq_kt})
simplify to
\begin{eqnarray}
\sqrt{\langle j_{\perp}^2\rangle}                  & = &
\sqrt{\pi}\cdot\langle |j_{\perp y}|\rangle \,\,\,\approx\,\,\,
\langle p_{\perp}\rangle\cdot \sigma_N \nonumber\\
\sqrt{\langle k_{\perp}^2\rangle}                  & = &
\sqrt{\pi}\cdot\langle | k_{\perp y}|\rangle \,\,\, \approx\,\,\,
\langle p_{\perp}\rangle\cdot\sqrt{\sigma_F^2-\sigma_N^2}
\label{eq_limit}
\end{eqnarray}
\bgc
\begin{figure}[h]
\parbox{6.5cm}{
     \includegraphics[width=7cm,height=5cm,
bbllx=0,bblly=0,bburx=550,bbury=410,clip=]{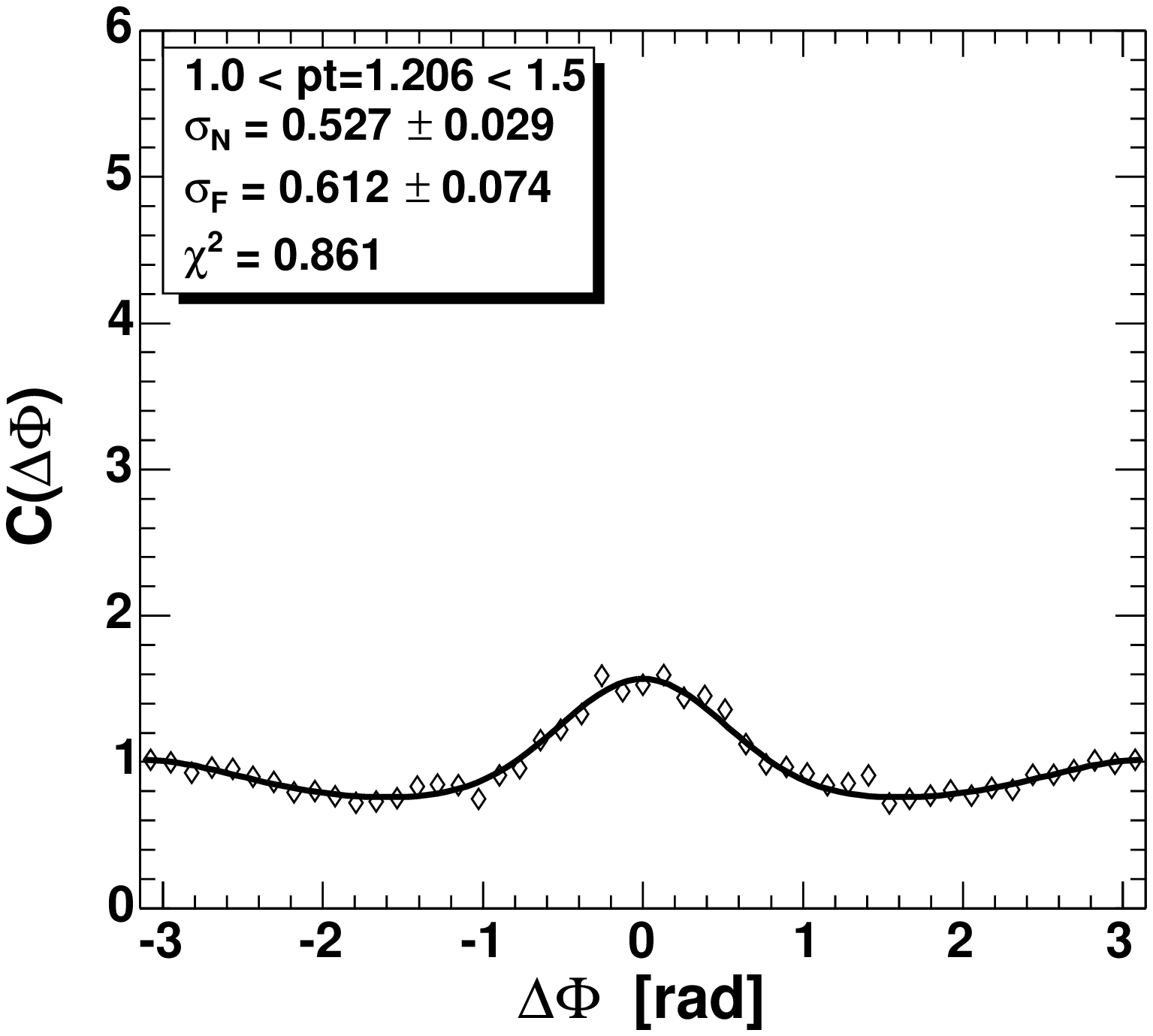}
     
} \hfill
\parbox{6.5cm}{
     \includegraphics[width=7cm,height=5cm,
bbllx=0,bblly=0,bburx=550,bbury=410,clip=]{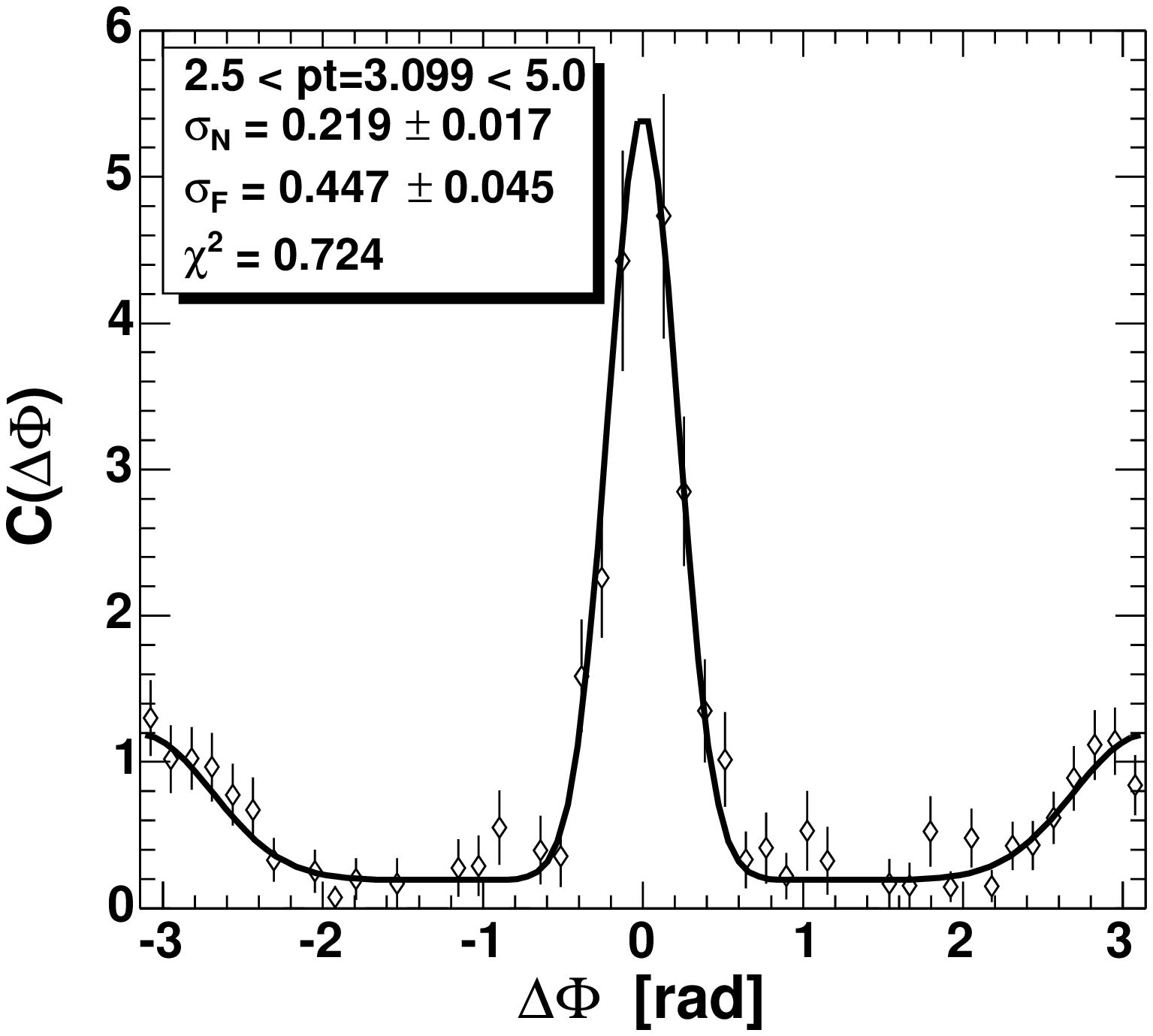}
      }
\caption{ 
Two charged hadrons area-normalized correlation function extracted
from $p+p$ data for both particles in 1.0$<p_\perp<$1.5 \gevc\ (left
panel) and 2.5$<p_\perp<$5.0 \gevc\ (right panel).  The solid line
correspond to the result of the fit (\ref{ppfit}).
}
\label{ppcorrelfce}
\end{figure}
\endc
\newpage\noindent
An example of the two-particle correlation function ($CF$), defined as
a ratio of ``real'' and ``mixed'' particle pairs distributions
$C(\Delta\phi)={dN_{real}\over d\Delta\phi}/{dN_{mixed}\over
d\Delta\phi}$, in $p+p$ collisions is shown in Fig.~\ref{ppcorrelfce}
for $\langle p_\perp\rangle$ = 1.2\gevc\ (left panel) and $\langle
p_\perp\rangle$ = 3.1\gevc\ (right panel). 
\bgc
\begin{figure}[h]
     \includegraphics[width=13cm,height=11cm]{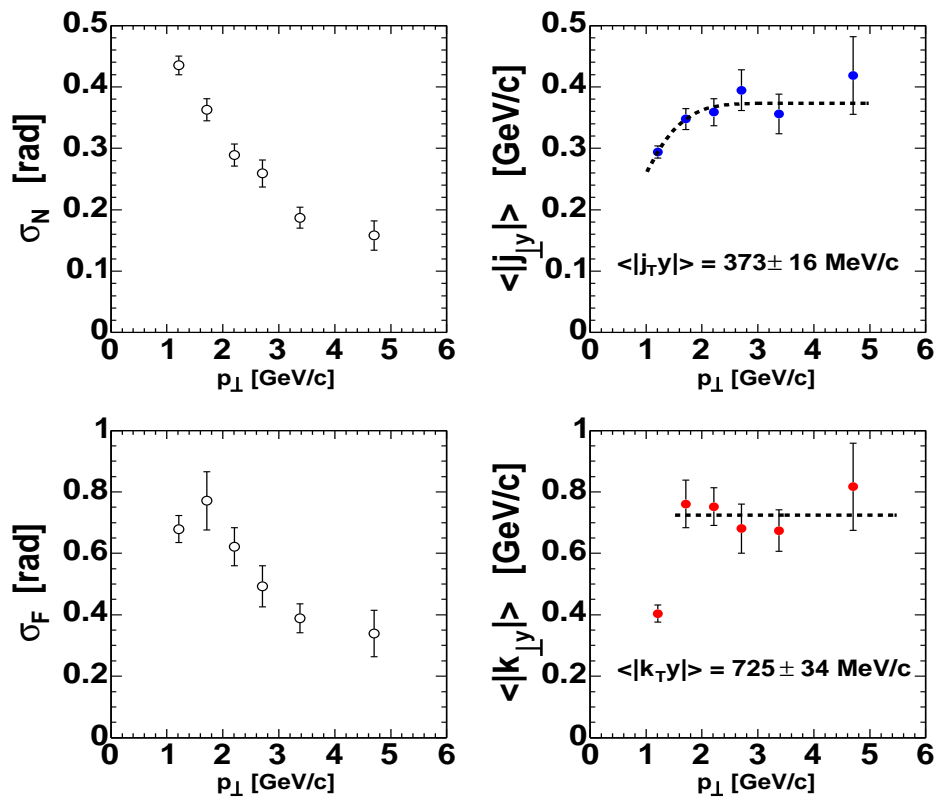}
\caption{ 
Measured width of the near angle peak $\sigma_N$ (upper left panel)
and far angle peak $\sigma_F$ (lower left panel) as a function of
\pt. Extracted values of \mjTy\ at given \pt-bins are shown on upper right panel.
The dashed line corresponds to the truncated ``Seagull'' fit (see
text).  The lower right panel shows the \mkTy\ with zeroth order
polynomial fit (dashed line).}
\label{jtkt}
\end{figure}
\endc
The correlation functions
are area-normalized ($\int_{-\pi}^{\pi}C(\Delta\phi)=2\pi$). A
significant excess of small angle ($\Delta\phi$=0, intra-jet
correlations) and large angle ($\Delta\phi=\pm\pi$, inter-jet
correlations) is evident. The data were fitted by
\bge\label{ppfit}
Fit(C,Y_N,\sigma_N,Y_F,\sigma_F) = C+Y_N\cdot Gauss(\Delta\phi=0,\sigma_N)+Y_F\cdot Gauss(\Delta\phi=\pm\pi,\sigma_F)
\ende
function, where $C$ is the constant, $Y_N$, $\sigma_N$, $Y_F$ and
$\sigma_F$ are the yields and peak widths of near and far angle peaks respectively.

The analysis of the correlation functions for various \pt-bins were
performed. Extracted values of $\sigma_N$ and $\sigma_F$ are shown on
Fig.~\ref{jtkt}. The angular width of the near and far angle peak is
decreasing with \pt\ as expected in case of jet fragmentation.  The
values of \mjTy\ and \mkTy\ were extracted. In case of \mjTy\ (upper
right panel of Fig.~\ref{jtkt}) one can observe the reduction of
\mjTy\ extracted for lowest \pt-bins. This is caused by \pt\
truncation, sometimes referred as ``Seagull effect'' \cite{Seagull}
(the \jT\ at given \pt\ cannot be larger then \pt\ itself, and thus
the measured $\sigma_N$ and \jT\ at $p_\perp\approx j_\perp$ is
accordingly reduced.)  One can easily account for such truncation and
the result of the fit procedure is shown as a dashed line on upper
right panel of Fig.~\ref{jtkt}. 
\bgc
\begin{figure}[h]
\parbox{6.5cm}{
     \includegraphics[width=7cm,height=7cm]{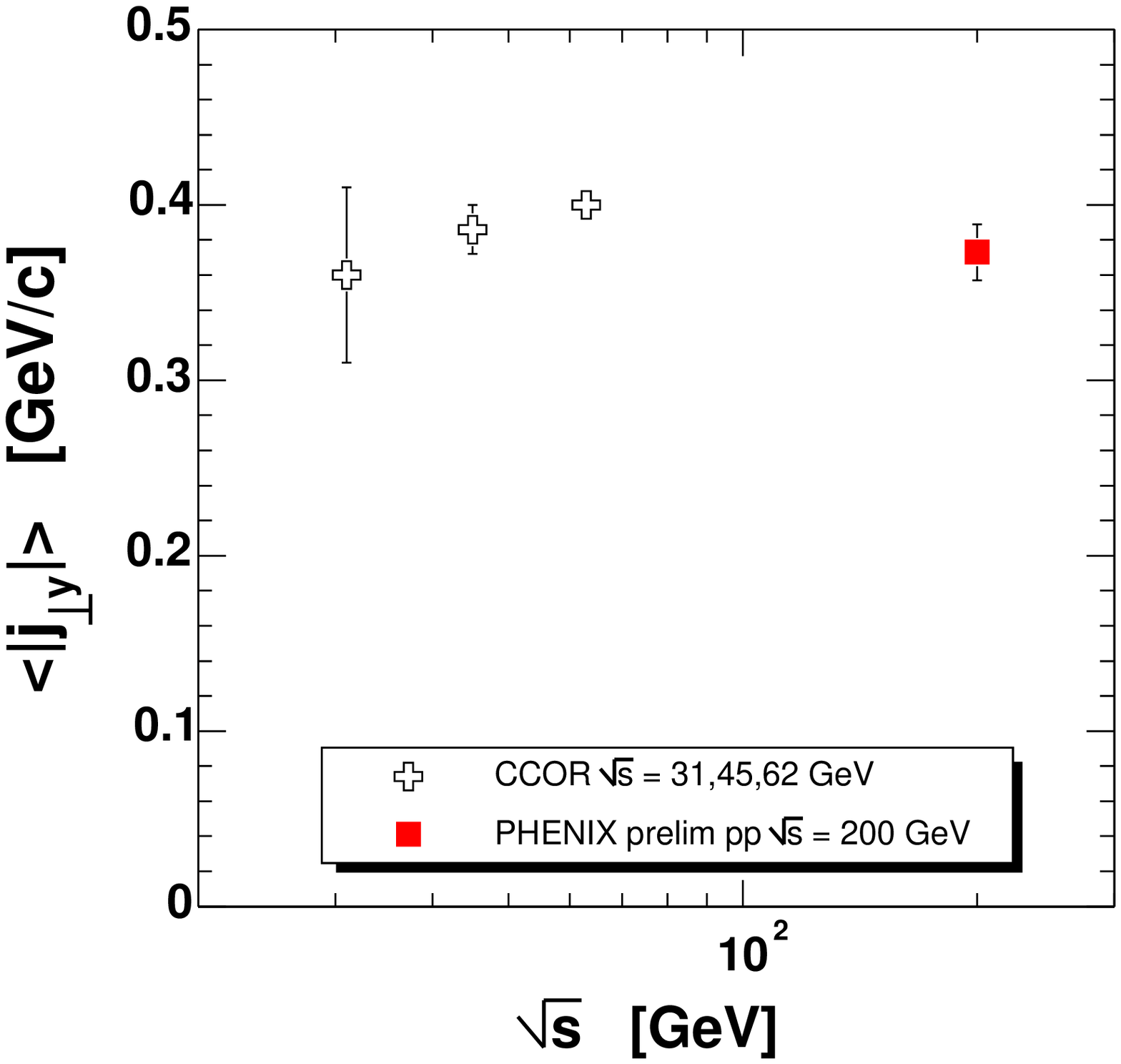}
     
} \hfill
\parbox{6.5cm}{
     \includegraphics[width=7cm,height=7cm]{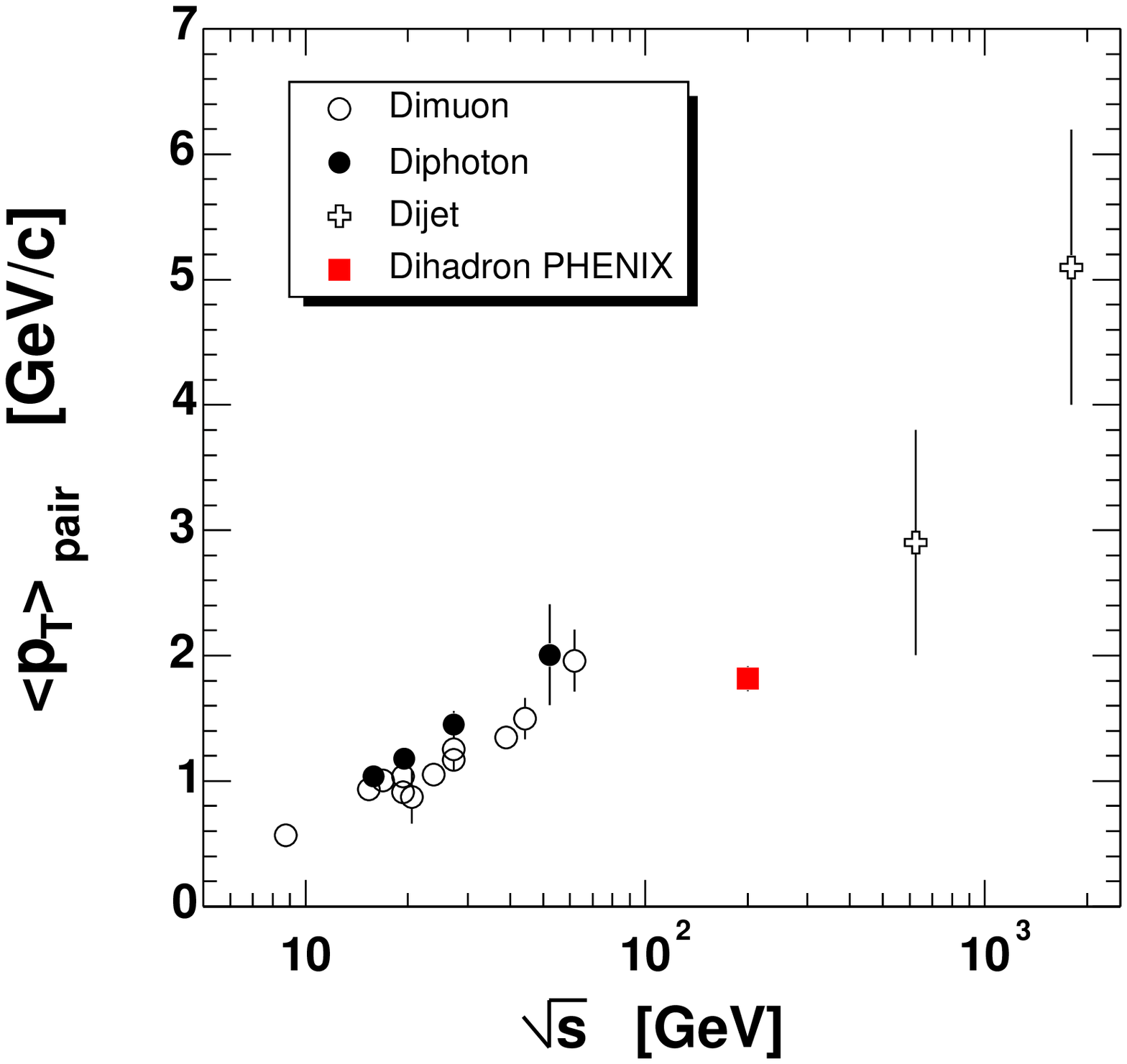}
      }
\caption{
The comparison of PHENIX preliminary measurement of \mjTy\ at \roots =
200\gev\ with CCOR measurement done at \roots = 31,45 and 62 \gev\
(left panel). The comparison of $\sqrt{\langle
p_{\perp}^2\rangle}_{pair}$ extracted from PHENIX preliminary
measurement on \mkTy\ according (\ref{ktpair}) shown on right panel.}
\label{CCORjtkt}
\end{figure}
\endc
\vskip -0.5 cm
The average value is \bgc\mjTy\ = 373 $\pm$ 16MeV/c.\endc
The average \mkTy\ was extracted by fitting a constant to the data
above 1.5\gevc. The average value found is
\bgc\mkTy\ = 725 $\pm$ 34 MeV/c. \endc
The comparison of \mjTy\ with results of CCOR collaboration
\cite{CCORjt} (left panel on Fig.~\ref{CCORjtkt}) shows a good
agreement. The \mkTy\ parameter could be compared to average RMS of the 
pair's momentum $\sqrt{\langle p_{\perp}^2\rangle}_{pair}$, which has
been studied on $p+\bar{p}$ Tevatron collider at Fermilab
\cite{Apana_kt}. The pair's \pt\ is related to the individual parton
\mkTy\ by the expression
\bge\label{ktpair}
\sqrt{\langle p_{\perp}^2\rangle}_{pair} =
\sqrt{2}\cdot\sqrt{\langle k_{\perp}^2\rangle} =
\sqrt{2\pi}\cdot\langle |k_{\perp y}|\rangle
\ende
Extracted value of $\sqrt{\langle
p_{\perp}^2\rangle}_{pair}$=1.82$\pm$0.85\gevc\ is in good agreement
with $p+\bar{p}$ data. This value could be also compared to $\langle
p_{\perp}\rangle$=1.80 $\pm$ 0.23 (stat) $\pm$ 0.16 (sys) \gevc\ of PHENIX
$J/\Psi$ measurement \cite{ppg017}.

\section{Two-particle correlation in $AA$}

Since the colliding heavy nuclei have a finite radius, the collision
zone, except in very central collisions, has typical ``almond'' like shape
with transverse dimensions of order of few fm. This geometrical
azimuthal asymmetry is usually characterized by eccentricity
\bge
\epsilon(b) = {<y^2-x^2>\over<y^2+x^2>}
\ende
where $x$ and $y$ are the coordinates of colliding nuclear
constituents in transverse plane and $b$ is the magnitude of the
impact parameter. In the hydrodynamical approach
\cite{hydro}, the original geometrical asymmetry is transformed into
momentum space by means of multiple interactions between particles. The
anisotropy in the momentum space is parametrized in the same way as
spatial anisotropy
\bge
v_2(p_\perp,b) = {<p_x^2-p_y^2>\over<p_x^2+p_y^2>}
\label{v2}
\ende
where $v_2(p_\perp,b)$ corresponds to the second harmonic
coefficient of Fourier expansion of the particles azimuthal distribution
\bge
{dN_f\over d\phi} \propto \left[ 1+2v_2\cdot\cos(2(\phi-\Psi))\right]
\label{dnf_dphi}
\ende
Here $\phi$ is the relative azimuthal angle of the particle emission with
respect to the reaction plane $\Psi$. In the case of ``pure'' flow,
when the particles' correlation is completely generated by
(\ref{dnf_dphi}), the correlation function has a form
\bge
{dN_f\over d\Delta\phi}=\int_{-\pi}^{\pi} d\phi {dN_f\over d\phi}\cdot{dN_f\over d(\phi+\Delta\phi)} 
\propto (1+2v^2_2\cdot\cos(2\Delta\phi))
\label{dnf_ddeltaphi}
\ende

It is obvious that the azimuthal anisotropy \vv\ generated by nuclear
geometry cannot exceed the magnitude of eccentricity $\epsilon$ at
given impact parameter. 

\begin{figure}[h]
\parbox{6.5cm}{\includegraphics[width=7cm,height=7cm]{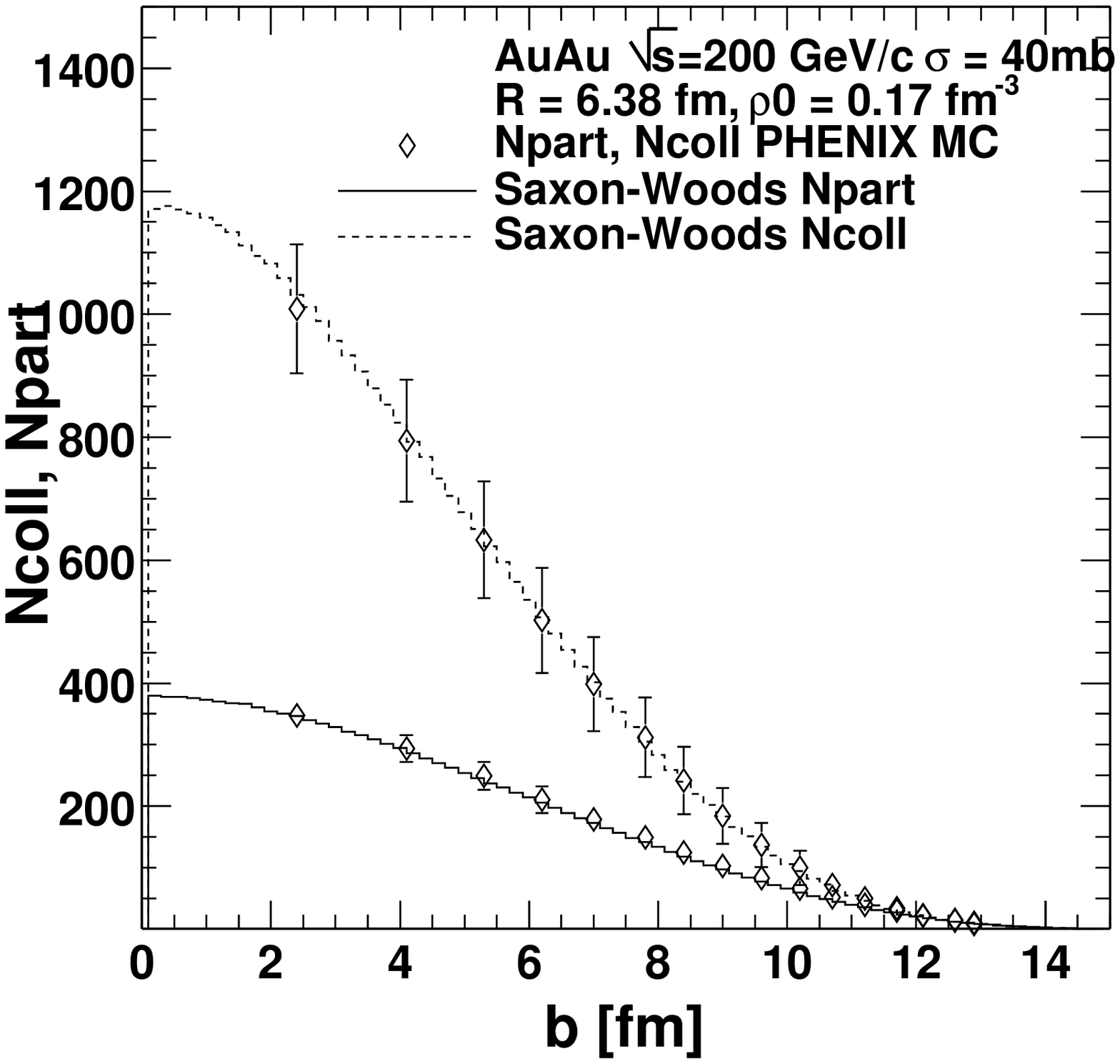}}\hfill
\parbox{6.5cm}{\includegraphics[width=7cm,height=7cm]{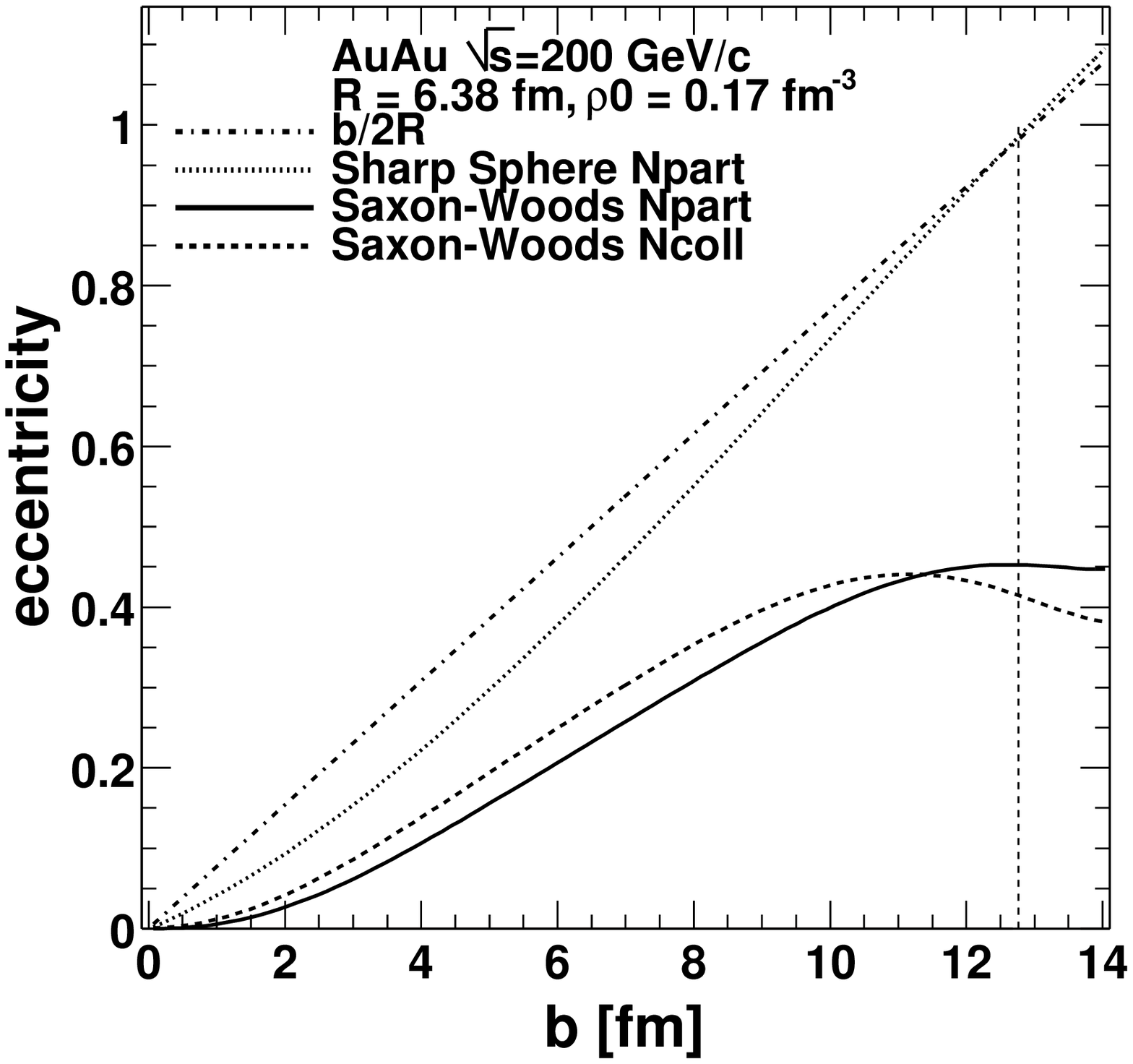}}
\caption{
Left panel: the $N_{coll}$ and $N_{part}$ distribution with impact
parameter $b$ computed for $Au+Au$ collisions at
$\sqrt{s_{NN}}$=200\gev. Total inelastic cross section
$\sigma_{tot}$=40mb, $Au$ radius $R$=6.38fm and nuclear density of
0.17 \gev/fm$^{3}$ were assumed.
Right panel: The eccentricity
dependence on $b$ for $N_{coll}$ and $N_{part}$ distributions
calculated with Saxon-Woods and sharp sphere nuclear density
distributions. The geometrical eccentricity $b/2R$ shown with
dashed-dotted line.
}
\label{eccentr}
\end{figure}

In the ``sharp sphere'' approach, where the nuclear density is
assumed to be constant inside and zero outside sphere of radius $R$,
the eccentricity of a collision at impact parameter $b=|\vec{b}|$ is
$\epsilon = b/2R$, and the geometrical \vv\ could be as large as 1; but
this approach is too idealized. When the more realistic nuclear density
distribution of ``Saxon-Woods'' type is taken into account
\cite{Glauber}, the maximum value of $\epsilon$ barely reaches 50\% (see Fig.\ref{eccentr}).
\bgc
\endc
However, the momentum anisotropy generated by collision geometry,
irrespective of details of collision dynamics, can be only
smaller. Since the anisotropy in the high-\pt\ region cannot be
generated by hydrodynamic flow (we will ignore the initial state
effects described in \cite{Kovchegov02}), the expected source of
observed anisotropy is the process of energy loss of high energy
partons in excited nuclear medium. The upper limit on \vv\ generated
by jet-quenching mechanism can be computed in a similar way described
in \cite{Shuryak}. 

The parton born in hard scattering events inside the collision almond
has to travel the distance $L$ in the medium before it reaches the
outer space. The ``primordial'' azimuthal asymmetry of the medium,
where the hard scattering event happened, gradually vanishes as the
medium expands. Since we are searching for an upper limit on \vv\
generated by jet-quenching mechanism we can neglect any dynamic and
expansion of the collision zone.

\begin{figure}[h]
\bgc
\includegraphics[width=8cm,height=7.0cm,
bbllx=0,bblly=100,bburx=550,bbury=520,clip=]{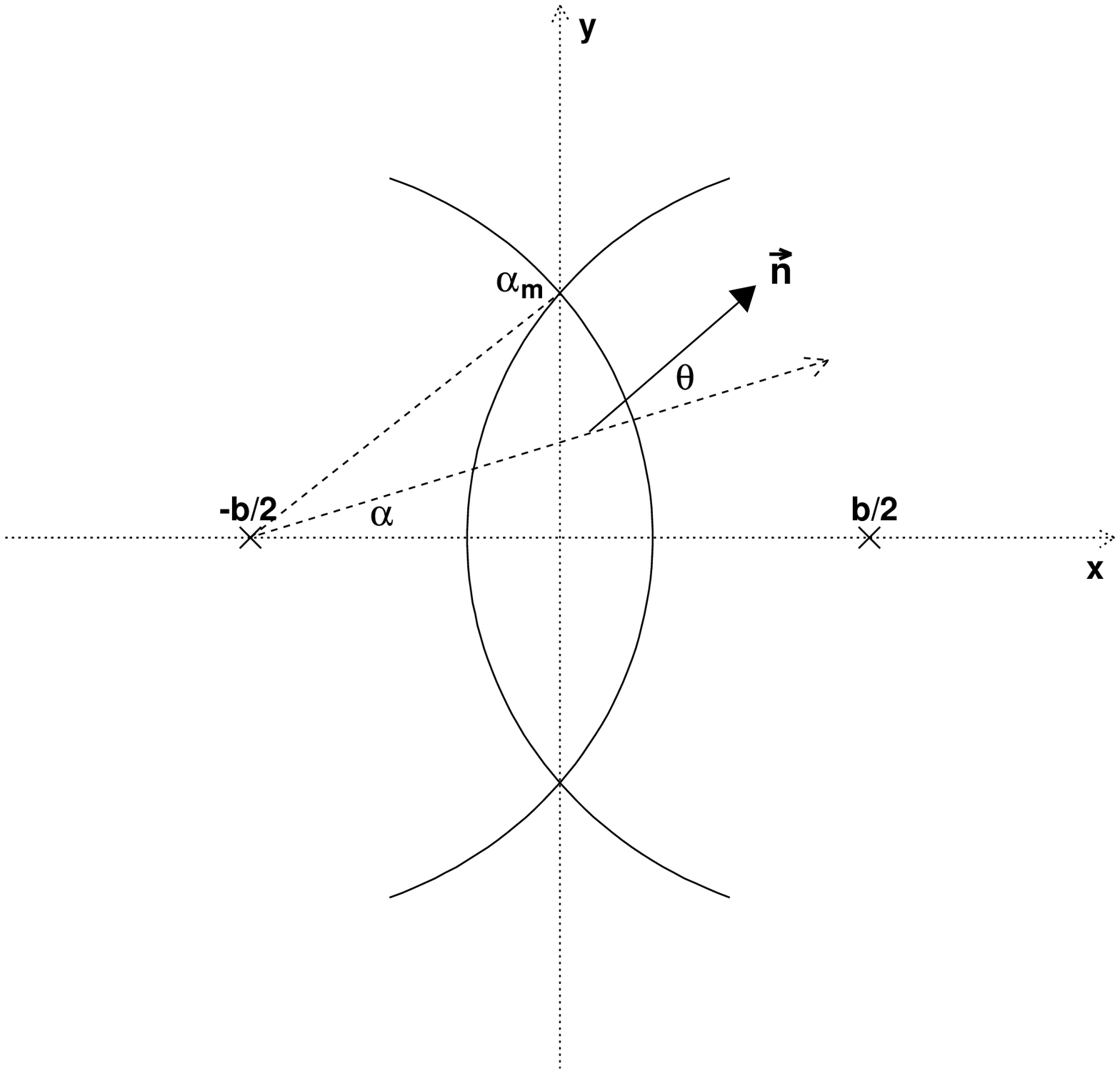}
\endc
\caption{
The schematic view of particle production in nuclear
collisions. Particle produced inside the overlap region propagates in
the direction $\vec{n}$.}
\label{schem}
\end{figure}

The energy loss of the parton propagating through the medium could be
characterized by absorption coefficient $\kappa$ (dimension
fm$^{-3}$). The probability of escaping the medium for parton born at
the space point $\vec{x}_0$, which propagates along the direction
$\vec{n}$, could be expressed as
\bge
f(\vec{x}_0,\vec{n})=exp[- \kappa \int_0^\infty ds\, L_-(\vec{x}_0+s\vec{n})\cdot L_+(\vec{x}_0+s\vec{n})\,].
\ende
(see Fig.~\ref{schem}). Here $L_\pm(x,y)=2[R^2-y^2-(x\pm b/2)^2]^{1/2}$, nuclear
thicknesses along the beam axis in case of sharp sphere approach,
characterizes the nuclear density at space point $(x,y)$ and it is
proportional to the probability of hard scattering \cite{Shuryak}.
\bgc
\begin{figure}[h]
\parbox{6.5cm}{
     \includegraphics[width=7cm,height=7cm]{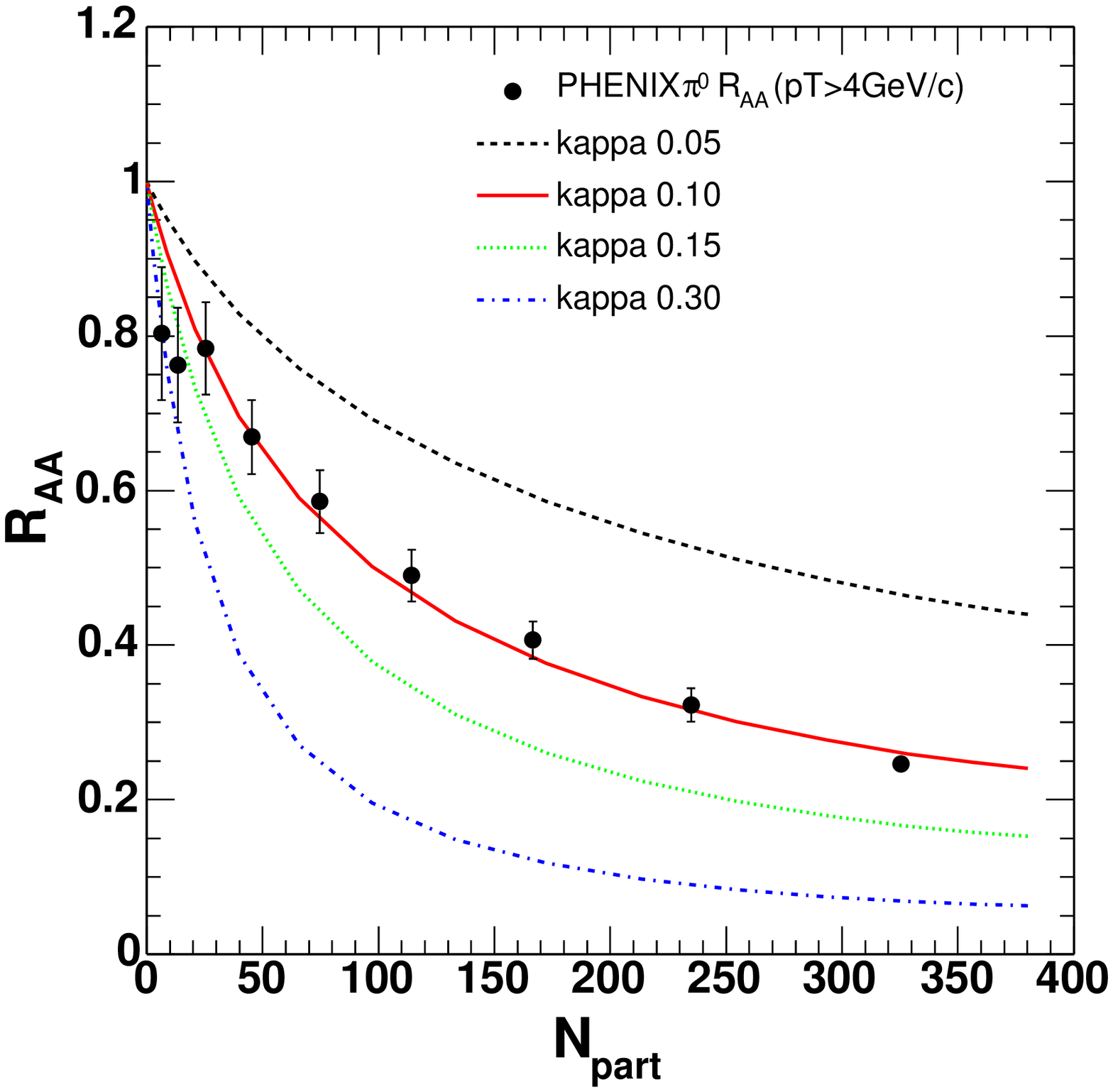}
     
} \hfill
\parbox{6.5cm}{
     \includegraphics[width=7cm,height=7cm]{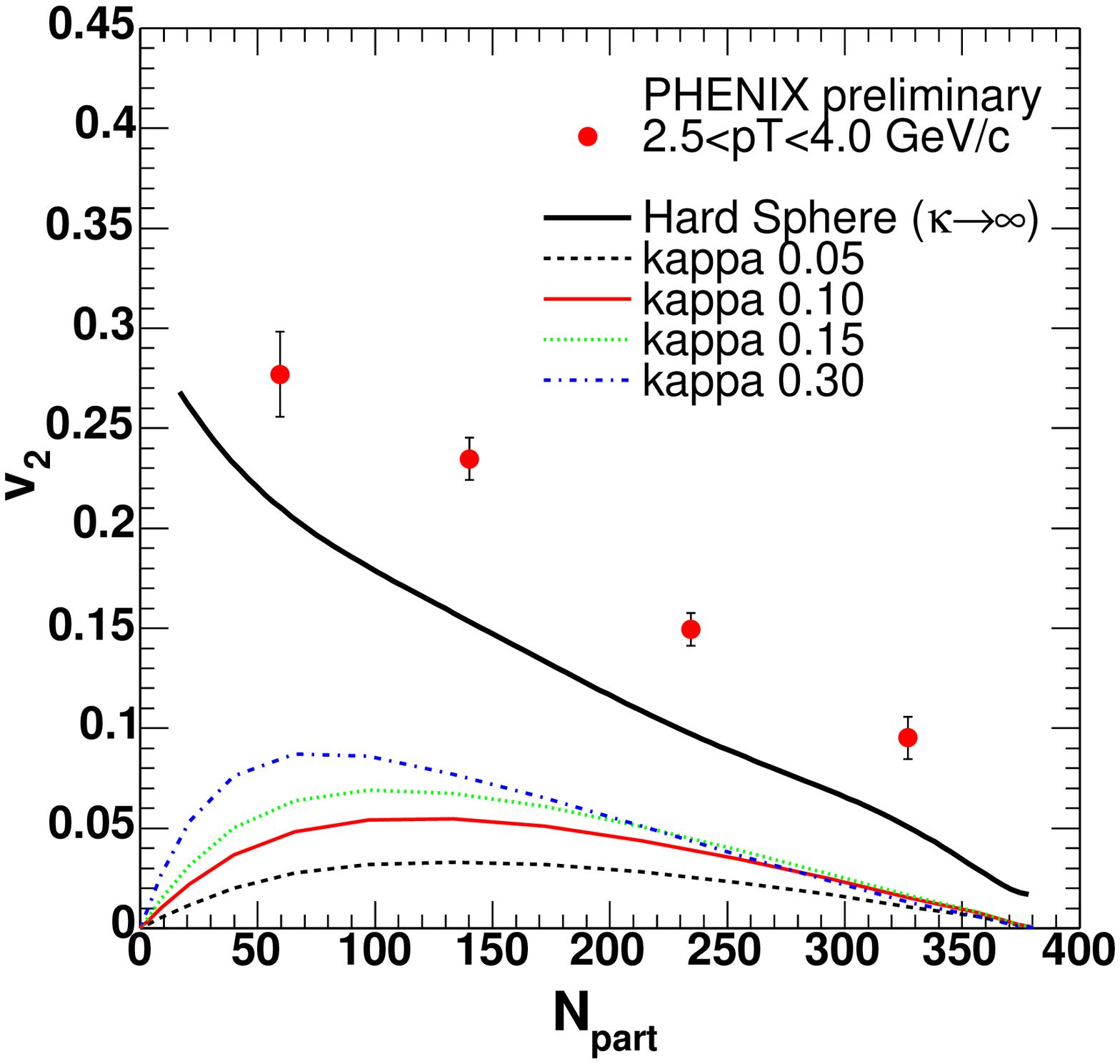}
      }
\caption{
Left panel: the relative absorption rates for various values of
parameter $\kappa$ compared to the PHENIX $\pi^0$ $R_{AA}$ for
$p_\perp>$4\gevc\ \cite{ppg014}. Right panel: calculated \vv\ as a
function of $N_{part}$ for the same values of parameter $\kappa$
compared with the PHENIX preliminary data \cite{ppg004}. The solid
line represents the limit from sharp sphere with
$\kappa\rightarrow\infty$ (see (\ref{voloshin})).  }
\label{kappa}
\end{figure}
\endc
In more realistic Glauber model one has to replace $L_-L_+$ by
$N_{part}(\vec{x})$ distribution, which characterizes the density of
excited nuclear medium.  The probability distribution of hard
scattering is approximated with the product of two thickness functions
$T_A(x,y)T_B(x,y)$ (see \cite{Glauber}).  As it was mentioned, this
approach provides only an upper limit of \vv\ from the jet quenching
mechanism. Any more realistic scenario with a detailed calculation of
partonic interactions with expanding QCD medium could lead only to
reduction of the resulting \vv.

The $\kappa$ parameter is the only unknown parameter in this
calculation and could be estimated from the comparison of the
absorption rate dependency on collision centrality expressed in terms
of $N_{part}$ (see Fig.~\ref{kappa}).  Adjusting the absorption
coefficient $\kappa$ comparing to PHENIX $\pi^0$ $R_{AA}$ for
$p_\perp>$4\gevc , we found the $\kappa$=0.1 fm$^{-3}$ provides the
best agreement with the data. But the maximum \vv\ in this case is
6\%. For less realistic values of $\kappa$ the maximum value of \vv\
reaches 10\%. However, the values of \vvpt\ measured by the PHENIX
collaboration using two-particle correlation techniques exceeds this
maximum ``jet-quenching'' \vvpt\ by a factor of 3.

The solid line on the right panel of Fig.~\ref{kappa} represents an idealized
limit in sharp sphere approach, when 
$\kappa\rightarrow\infty$, and only radiation emitted from the
surface is not absorbed \cite{Voloschin_QM}. The escape probability $\mathrm{P}_{esc}$ is
in the $\sigma\rightarrow\infty$ limit 
\bge
\mathrm{P}_{esc}  =  \int_{0}^{\infty}dh \, exp(-\rho\sigma h/\cos{\theta}) \,\,\,\propto\,\,\cos{\theta}
\arrowvert_{\sigma\rightarrow\infty}.
\ende
where $\rho$ stands for nuclear density, $\sigma$ for absorption cross
section and the product $\rho\sigma$ has similar meaning as $\kappa$ parameter.
The \vv\ can be obtained from integration
\bge\label{voloshin}
v_2  =  \langle  cos(2\phi)\rangle = \int d\phi \,\mathrm{P}_{esc}\cdot cos(2\phi)
= \int_{-\alpha_m}^{\alpha_m}\!\!\!\!\!\! d\alpha\int_{-\pi/2}^{\pi/2}\!\!\! d\theta\cos{\theta}
{\cos{2(\theta+\alpha)}\over 2\cdot 2\alpha}
= {\sin{2\alpha_m}\over 6\alpha_m}
\ende
where $\alpha_m=\sqrt{R^2-(b/2)^2}$.

As one can see on Fig.~\ref{kappa}, even the unphysical model of black body radiation
providing the upper ``geometrical'' limit on \vv\ , still does not
generate enough azimuthal anisotropy as observed in data. It is
evident, that in order to to explain measured azimuthal anisotropies,
some new underlying mechanism, other than just the ``jet-quenching'', has
to be invoked.

\section{Summary}

We have presented the detailed shape analysis of the two-hadron
correlation function in $p+p$ collisions at \roots=200\gev. The
average jet-transverse momentum \mjTy\ = 373 $\pm$ 16MeV/c was
measured.  The average partons transverse momentum \mkTy\ was
extracted by fitting the constant to the data above 1.5\gevc, and the
average value of \mkTy\ = 725 $\pm$ 34 MeV/c was found.  The
comparison of \mjTy\ with measurement done by CCOR experiment
\cite{CCORjt} shows good agreement and no significant trend with
increasing \roots. The value of \mkTy\ was compared to the values
extracted from data take at Tevatron collider, and no significant
deviation from the overall trend was found. The indirect measure of
partonic primordial transverse momentum is the average \pt\ of
$J/\Psi$ mesons. Such a measurement has been done by PHENIX
collaboration as well, and the value of $\langle
p_{\perp}\rangle$=1.80 $\pm$ 0.23 (stat) $\pm$ 0.16 (sys) \gevc\ has
been found to be in good agreement with the $\sqrt{\langle
p_{\perp}^2\rangle}_{pair}$ = 1.82 $\pm$ 0.09 \gevc\ (see eq. \ref{ktpair}).

The analysis of two-particle correlation in $Au+Au$ at \roots =
200\gev\ is also presented. Using the simple Glauber model we have
demonstrated that the elliptic flow parameter \vv\ cannot be generated
solely by jet-quenching mechanism. The upper limit for \vv\ from
jet-quenching has been found to be less than 10\%. It is evident that
the large values of \vv\ observed by all RHIC experiments have to be
generated by other mechanisms. There is at least one model, inspired
by ``Color Glass Condensate'' \cite{Kovchegov02}, which suggests the
two-gluon correlations as a source of large azimuthal
anisotropies. However; the elliptic anisotropy, in this model, does
not respect the reaction plane. This seems to be in contradiction with
observed correlation between the reaction plane and spatial
distribution of particle emission after freeze-out observed by STAR
collaboration \cite{STAR_HBT}. An interesting alternative provides
the ``coalescence'' model, where the partonic flow is ``amplified'' by
means of parton recombinations \cite{Reiner_Recombination}. This
scenario is very exciting since it requires some kind of partonic
collectivity in the initial state of the collision, which is
anticipated if the quark-gluon plasma is created.

\end{document}